\def\rfr#1{eq. (\ref{#1})}

\def\derp#1#2{\rp{\partial{#1}}{\partial{#2}}}

\def\bar{\begin{eqnarray}}
\def\ear{\end{eqnarray}}
\def\bb{\bibitem}
\def\eqi{\begin{equation}}
\def\eqf{\end{equation}}
\def\eqia{\begin{eqnarray}}
\def\eqfa{\end{eqnarray}}
\def\rp#1#2{{#1\over#2}}

\def\lb#1{\label{#1}}

\def\oc2{$\mathcal{O}(c^{-2})$}

\documentclass{aastex}
\usepackage{spr-astr-addons}
\usepackage{url}

\RequirePackage{color}

\newcommand{\emaila}{lorenzo.iorio@libero.it}

\begin{document}

\title{The impact of the oblateness of Regulus on the motion of its companion}

\shorttitle{The impact of the oblateness of Regulus on the motion of its companion}
\shortauthors{L. Iorio }

\author{Lorenzo Iorio\altaffilmark{1}}
\affil{INFN-Sezione di Pisa. Permanent address for correspondence: Viale Unit\`{a} di Italia 68, 70125, Bari (BA), Italy.}

\email{\emaila}

\begin{abstract}
  {The fast spinning B-star Regulus has recently been found to be orbited by a fainter companion in a close circular path with orbital period  $P_{\rm b} = 40.11(2)$ d. Being its equatorial radius $R_{\rm e}$ $32\%$ larger than the polar one $R_{\rm p}$, Regulus possesses a remarkable quadrupole mass moment $Q$.}
  {We investigate the effects of $Q$ on the orbital period $P_{\rm b}$ of its companion in order to see if they are measurable, given the present-day level of accuracy in measuring $P_{\rm b}$. Conversely, we will look for deviations from the third Kepler law, attributed to the quadrupole mass moment $Q$ of Regulus, to constrain the ratio $\gamma=m/M$ of the system's masses.}
  {The impact of $Q$ on the orbital period is analytically worked out with a straightforward perturbative approach. The resulting correction $P^Q$ is compared to other competing dynamical effects.
  $P^Q$ and the Keplerian period $P^{\rm Kep}$ are expressed in terms of the phenomenologically determined system's parameters; $\gamma$ is treated as an unknown. $P^Q$ is compared to the observational accuracy in measuring the orbital period $\delta P_{\rm b}=0.02$ d and to the systematic uncertainty  $\delta(P^{\rm Kep})$ due to the errors in the system's parameters entering it. The discrepancy $\Delta P=|P_{\rm b}-P^{\rm Kep}|$ is examined in order to see for which values of $\gamma$ it becomes statistically significant. The physical meaning of the obtained range of values for $\gamma$ is discussed in terms of $Q$.}
  {$P^Q$ is larger than $\delta P_{\rm b}$ but still smaller than the systematic uncertainty in $P^{\rm Kep}$ by two orders of magnitude. The major sources of bias are the velocity semiamplitude $K$ of the motion of the primary and its mass $M$. Assuming edge-on configuration, i.e. $i=90$ deg, if $\gamma\gtrsim 0.096$ $Q$ would be positive, i.e. Regulus would be prolate, contrary to the observations. If $\gamma\lesssim 0.078$ $Q$ would be negative, but its magnitude would be one-two orders of magnitude larger than the approximate estimate $Q\approx M(R^2_{\rm p}-R^2_{\rm e})=-2.4\pm 0.5\times 10^{49}\ {\rm kg\ m}^2$. }
  {Regulus is the first extrasolar binary system in which the orbital effects of the asphericity of the primary are larger than the observational sensitivity; moreover, no other competing aliasing orbital effects are present. Thus, it is desirable that it will become the object of future intensive observational campaigns in order to reduce the systematic uncertainty due to the system's parameters below the measurability threshold.}

\end{abstract}


\section{Introduction}
Regulus ($\alpha$ Leo, HR 3982, HD 87901) is a nearby ($d=24.3\pm 0.2$ pc \citep{vanL07}) intermediate mass star of spectral class B \citep{Joh53,Gr03} and has the peculiarity of being animated by a very fast rotation. Indeed, combined interferometric and spectroscopic studies \citep{McAL05} have shown that its equatorial radius $R_{\rm e}$ is $32\%$ larger than the polar one $R_{\rm p}$, and its rotation period is 15.9 hr, which corresponds to an equatorial rotation speed that is $86\%$ of the critical breakup velocity. Other fast spinning stars whose oblateness has been interferometrically measured  are Altair ($\alpha$ Aquil{\ae}, HD 187642 \citep{Alt}) and Achernar ($\alpha$ Eri, HD 10144 \citep{Acher}).

In principle, a companion orbiting not too far from such  highly deformed stars would experience relevant orbital effects induced by the primary's oblateness which could be used  to dynamically put constraints either on such an important stellar physical parameter or on the orbital/physical parameters of the system.
In fact, the Be star Achernar recently turned out to be orbited by a faint companion \citep{Acherdoppia} which should be a A1V-A3V star \citep{AcherB}, but since its orbital period should amount to about 15 yr it has not yet been possible to obtain its orbital parameters. No companions are known for Altair.
The situation is more favorable for Regulus since spectroscopic observations over the last few years have demonstrated that it has a close\footnote{In fact, Regulus has a known wide companion $\alpha$ Leo B at a separation of $\approx 175''$ which is itself a binary \citep{McAL05}, but it has a far too great separation to have ever interacted directly with Regulus. } companion in a circular orbit described in 40.11 d \citep{Gies08}.

In this paper we will investigate the possibility of detecting some dynamical orbital effects induced by the oblateness of Regulus on its companion and will use them to put some constraints on the mass of the secondary.


\section{The impact of $Q$ on the orbital period and its measurability}
The relevant physical and orbital parameters of the Regulus binary system are in Table \ref{tavola}.
\begin{table}[t]
\caption{ Relevant parameters of the single-line spectroscopic binary system of Regulus. $M$ is the primary's mass \citep{McAL05}, $P_{\rm b}$ is the orbital period \citep{Gies08}, $K$ is the velocity semiamplitude \citep{Gies08}, $R_{\rm e}$ and $R_{\rm p}$ are the equatorial and polar radii \citep{McAL05}; the eccentricity $e$ has been assumed equal to zero \citep{Gies08}. Numbers in parentheses give the error in the last digit quoted.}
\label{tavola}
\begin{tabular}{@{}lllll}
\hline
$M$ (M$_{\odot}$) & $P_{\rm b}$ (d) & $K$ $\left(\rp{\rm km}{\rm s}\right)$  & $R_{\rm e}$ (R$_{\odot}$) & $R_{\rm p}$ (R$_{\odot}$)\\
\tableline
$3.4(2)$ & $40.11(2)$ & $7.7(3)$  & $4.16(8)$ & $3.14(6)$\\
\hline
\end{tabular}
\end{table}
Concerning the inclination $i$ to the plane of the sky, \citet{McAL05} showed that the best results for their fits are obtained by choosing  $i=90$ deg; in the following we will keep it fixed to different values close to the edge-on configuration. The estimates of stellar radii and mass are almost independent of the choice of $i$ \citep{McAL05}.

Given that the observable quantity at our disposal is the orbital period, we will start by investigating the impact of the primary's oblateness on it.
By assuming axial symmetry about the $z$ axis and reflection symmetry
about the equator assumed as reference $\{xy\}$ plane,
the  external
gravitational potential $U$ can be written as \citep{LAR}\eqi U=
U_0+U_{\mathcal{Q}},\eqf
\begin{equation}\left\{\begin{array}{lll}
U_0=-\rp{G{\mathcal{M}}}{r},\\\\
U_{\mathcal{Q}}=-\rp{G{\mathcal{Q}}}{r^3}\left(\rp{3\cos^2\theta-1}{2}\right),\lb{Q}
\end{array}\right.\end{equation}
in which $\theta$ is the colatitude angle and $\mathcal{Q}\equiv Q + q\approx Q$; indeed, the quadrupole mass moment is proportional
to the square of the stellar angular rotation frequency, so that we will neglect the oblateness $q$, if any, of the secondary with respect to that of Regulus.
The relative acceleration $\mathit{\mathbf{A}} = -\nabla U$ due to the gravitational potential of
\rfr{Q} is, in spherical coordinates
\begin{equation}\left\{\begin{array}{lll}
A_r = -\derp U r,\\\\
A_{\theta} = -\rp{1}{r}\derp U\theta,\\\\
A_{\phi} = -\rp{1}{r\sin\theta}\derp U\phi,
\end{array}\right.\end{equation}
 which yields
\begin{equation}\left\{\begin{array}{lll}
A_r=-\rp{G{\mathcal{M}}}{r^2}-\rp{3}{2}\rp{GQ}{r^4}(3\cos^2\theta-1),\\\\
A_{\theta}=-6\rp{GQ}{r^4}\sin 2\theta,\\\\
A_{\varphi}=0.\lb{Acci}
\end{array}\right.\end{equation}

We will now make the simplifying assumption that the orbital
angular momentum and the spin of Regulus are aligned \citep{McAL05,Gies08},
 Thus, $A_\theta=A_\varphi=0$ and only the equation for the
radial acceleration survives  in \rfr{Acci} as \eqi A_r =
A_0+A_Q,\eqf with
\begin{equation}\left\{\begin{array}{lll}
A_0=-\rp{G\mathcal{M}}{r^2}\\\\
A_Q= \rp{3}{2}\rp{GQ}{r^4}.\lb{radialacc}
\end{array}\right.\end{equation}
Following straightforwardly a standard perturbative approach \citep{Ior08}
is it possible to use \rfr{radialacc} to obtain
\eqi P = P^{\rm Kep} + P^Q,\eqf
with \eqi P^{\rm Kep} = 2\pi\sqrt{\rp{a^3}{G\mathcal{M}}}=(1+\gamma )\sqrt{\rp{1}{2\pi GM}\left(\rp{KP_{\rm b}}{\gamma \sin i}\right)^3 },\lb{pkpl}\eqf and
\eqi P^Q = \rp{3\pi Q}{\sqrt{Ga{\mathcal{M}}^3}}=\rp{3Q}{(1+\gamma )}\sqrt{\rp{2\gamma \sin i}{GKP_{\rm b}}\left(\rp{\pi}{M}\right)^3}.\lb{pqu}\eqf
We have used
\eqi \mathcal{M}=M+m=M(1+\gamma ),\ \gamma =\rp{m}{M},\eqf
and \eqi a = \left(1+\rp{M}{m}\right)a_M=\left(\rp{1+\gamma }{\gamma }\right)\rp{KP_{\rm b}}{2\pi\sin i},\eqf
where $a$ is the relative semimajor axis and $a_M$ is the barycentric semimajor axis of the motion of the primary.

\begin{figure}[t]
   \includegraphics[width=\columnwidth]{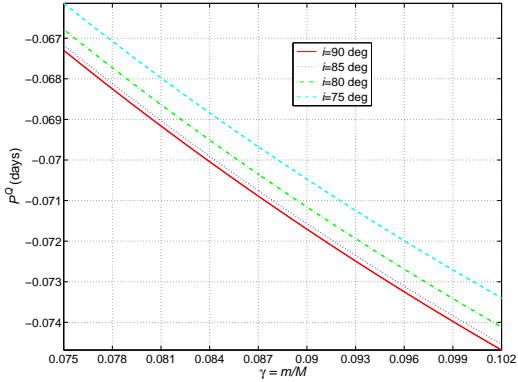}
   \caption{Quadrupole mass moment correction $P^{Q}$ to the Keplerian period for $M=3.4$M$_{\odot}$, $Q\approx M(R_{\rm p}^2-R_{\rm e}^2)=(-2.4\pm 0.5)\times 10^{49}$ kg m$^2$ and different values of the inclination $i$ close to the edge-on configuration. It is larger than the accuracy in the determined orbital period $\delta P_{\rm b}=0.02$ d.}
   \label{regulus_PQ}
   \end{figure}

Since  \citep{LAR}
\eqi Q=I_x - I_z\approx M(R_{\rm p}^2 - R_{\rm e}^2)=-2.4\pm 0.5\times 10^{49}\ {\rm kg\ m}^2, \lb{inerz}\eqf where $I_x$ and $I_z$ are the equatorial and polar moments of inertia, respectively,
\rfr{pqu} allows us to investigate if the contribution of the primary's oblateness to the orbital period is large enough to be detected, in principle, given the present-day level of accuracy in determining $P_{\rm b}$.  Figure \ref{regulus_PQ} and Table \ref{tavola} show that this is just the case. Here and in the following we will treat $\gamma $ as an unknown quantity.
This is an important result because it opens, in principle, the possibility of inferring or constraining $Q$ from the dynamics of the binary system. Moreover, it is the first time that such an opportunity occurs for an extrasolar binary system; indeed, we will show that other competing dynamical effects, which in other systems may overcome those due to $Q$, in this case are negligible.

\begin{figure}[t]
   \includegraphics[width=\columnwidth]{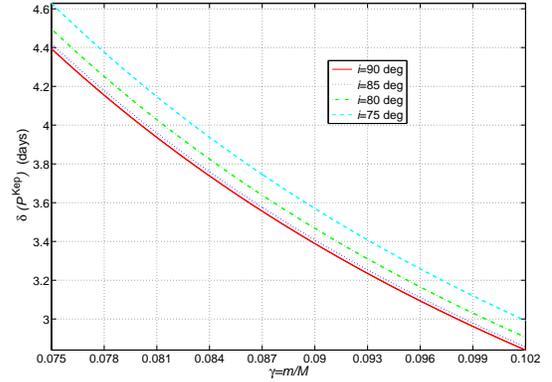}
   \caption{Total uncertainty $\delta (P^{\rm Kep})$ in the Keplerian period due to the errors in the system's parameters for $M=3.4$M$_{\odot}$ and different values of the inclination $i$ close to the edge-on configuration. It is larger than $P^Q$ by about two orders of magnitude.}
   \label{regulus_dP}
   \end{figure}

To do that it would be necessary compute the Keplerian orbital period $P^{\rm Kep}$ and subtract it from the measured one $P_{\rm b}$; to this aim, let us note that the parameters entering \rfr{pkpl} have been determined independently of the third Kepler law itself, so that it does make sense to compare \rfr{pkpl} to the phenomenologically determined orbital period $P_{\rm b}$.
 However, it is not sufficient that $\delta P_{\rm b}< P^Q$; is the uncertainty $\delta (P^{\rm Kep})$ due to the errors in the system's parameters smaller than $P^Q$ as well? Unfortunately, the answer is still negative by about two orders of magnitude, as shown by Figure \ref{regulus_dP}.

\begin{figure}[t]
   \includegraphics[width=\columnwidth]{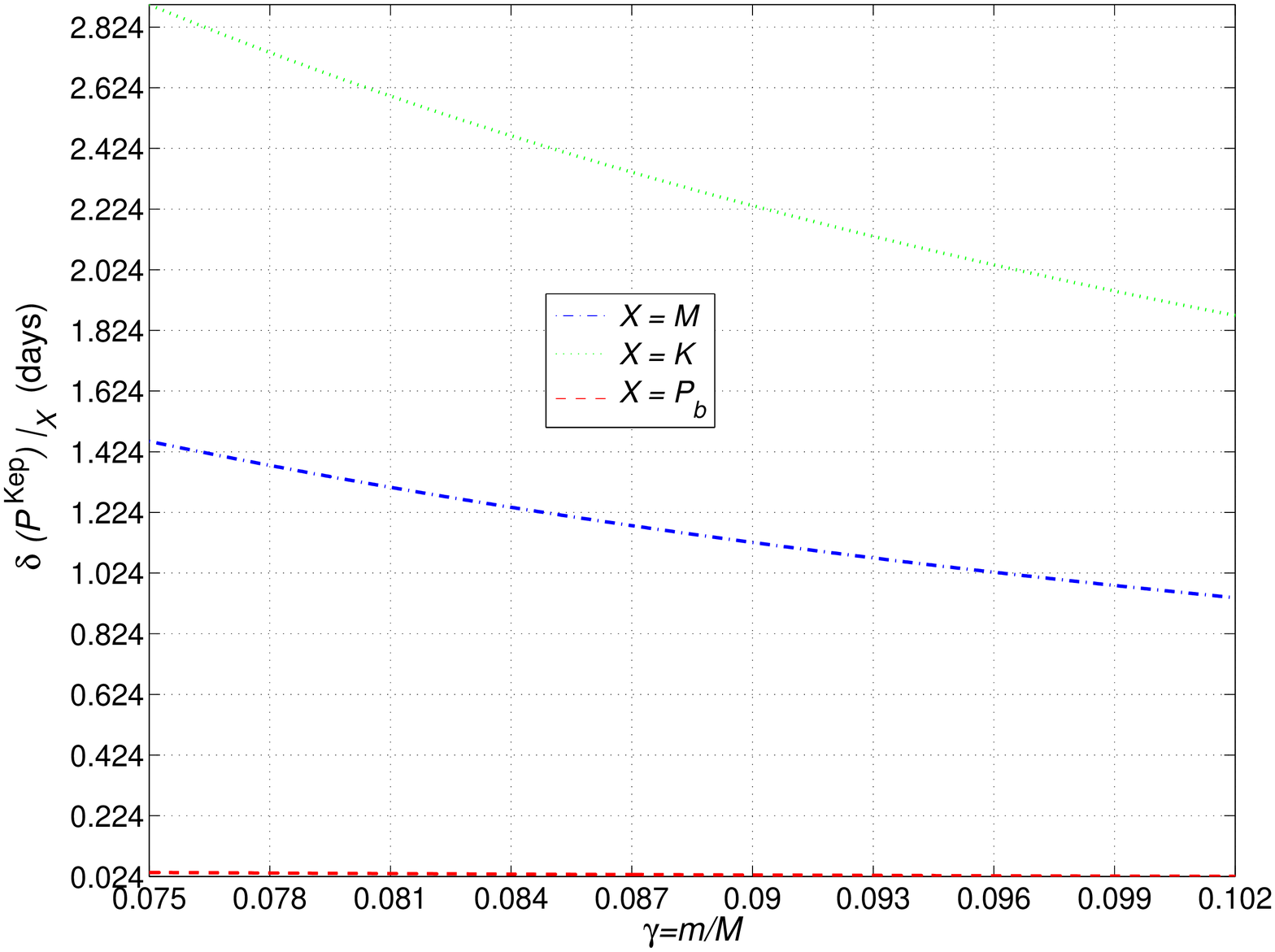}
   \caption{Uncertainties in the Keplerian period due to the errors in the system's parameters $M, K, P_{\rm b}$ for $M=3.4$M$_{\odot}$ and $i=90$ deg.}
   \label{regulus_dp}
   \end{figure}

It may be interesting to see what is the impact of each  system's parameter in determining $\delta(P^{\rm Kep})$ in order to have some hints about the future possibilities and also to drive researches towards the improvement of the parameters which turn out to be the most effective in corrupting the measurement of $P^Q$. Figure \ref{regulus_dp}
shows that  the most relevant sources of errors are $K$ and $M$.  For details concerning how $K$ and $M$ and their errors have been determined see \citep{Gies08} and references therein, and \citep{McAL05}, respectively.

In principle, there is also a general relativistic contribution to the orbital period which should be taken into account.
It is   \citep{Dam}
\eqi P^{\rm rel}\approx \rp{3\pi}{c^2}\sqrt{G\mathcal{M}a}=\rp{3}{c^2}(1+\gamma )\sqrt{\rp{\pi  GMKP_{\rm b}}{2\gamma \sin i}}.\eqf
However, it can be safely neglected because it is smaller than $\delta P_{\rm b}$ by four orders of magnitude, being of the order of $10^{-6}$ d.
%
%
%
%
%
%
%
%
%
%

Finally, let us note that, in view of the large separation between the two stars no tidal effects come into play.  This can be seen by comparing the tidal acceleration
\eqi A_{\rm tidal}=\rp{2GmR_{\rm e}}{a^3}=2GM\gamma^4 R_{\rm e}\left[\rp{2\pi\sin i}{(1+\gamma )K P_{\rm b}}\right]^3,\eqf
to the centrifugal acceleration
\eqi A_{\rm cen}= \Omega^2 R_{\rm e},\eqf where $\Omega$  is the rotational frequency of Regulus; 
it turns out that $A_{\rm tidal}/A_{\rm cen}\ll 1$ being of the order of $10^{-9}$.
%
%
%
%
%
%
%
%
%
%
\section{Constraints on the secondary's mass from deviations from the third Kepler law attributed to $Q$}
In this Section we will show how useful constraints on the mass $m$ of the Regulus' companion can be inferred by looking for deviations from the third Kepler law in terms of $Q$; indeed, in the previous Section we showed that no other effects are relevant in the orbital dynamics of such a binary system.

\begin{figure}[t]
   \includegraphics[width=\columnwidth]{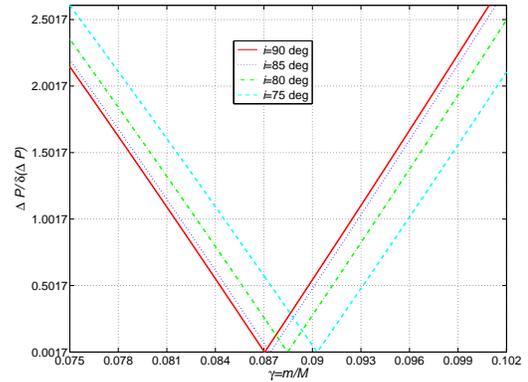}
   \caption{Ratio $\Delta P/\delta(\Delta P)$ of  the discrepancy $\Delta P=|P_{\rm b}-P^{\rm Kep}| $ between the measured orbital period $P_{\rm b}$ and the Keplerian one $P^{\rm Kep}$ to its uncertainty as a function of  $\gamma=m/M$  for $M=3.4$M$_{\odot}$ and different values of the inclination $i$ close to the edge-on configuration. It is significant at $1-\sigma$ level outside the range $0.081 \lesssim \gamma \lesssim 0.093$ ($i=90$ deg).}
   \label{regulus_P}
   \end{figure}

Let us start by evaluating the discrepancy $\Delta P=|P_{\rm b} - P^{\rm Kep}|$ between the measured orbital period $P_{\rm b}$ and the computed Keplerian one $P^{\rm Kep}$ as a function of $\gamma $ for different values of the inclination close to the edge-on configuration: Figure \ref{regulus_P} shows that, for $i=90$ deg, $\Delta P$ is significant at $1-\sigma$ level outside the range $0.081 \lesssim \gamma \lesssim 0.093$; for $i=75$ deg it occurs outside  $0.085 \lesssim \gamma \lesssim 0.096$.
    %
    %
Note that such bounds are conservative because the uncertainty in $\Delta P$ has been evaluated as $\delta(\Delta P)\leq \delta P_{\rm b} + \delta (P^{\rm Kep})$.
 For the sake of simplicity, let us consider the case $i=90$ deg and see  if statistically significant deviations from the third Kepler law for $\gamma\gtrsim 0.09$, interpreted as due to $Q$, are physically meaningful.
\begin{figure}[t]
   \includegraphics[width=\columnwidth]{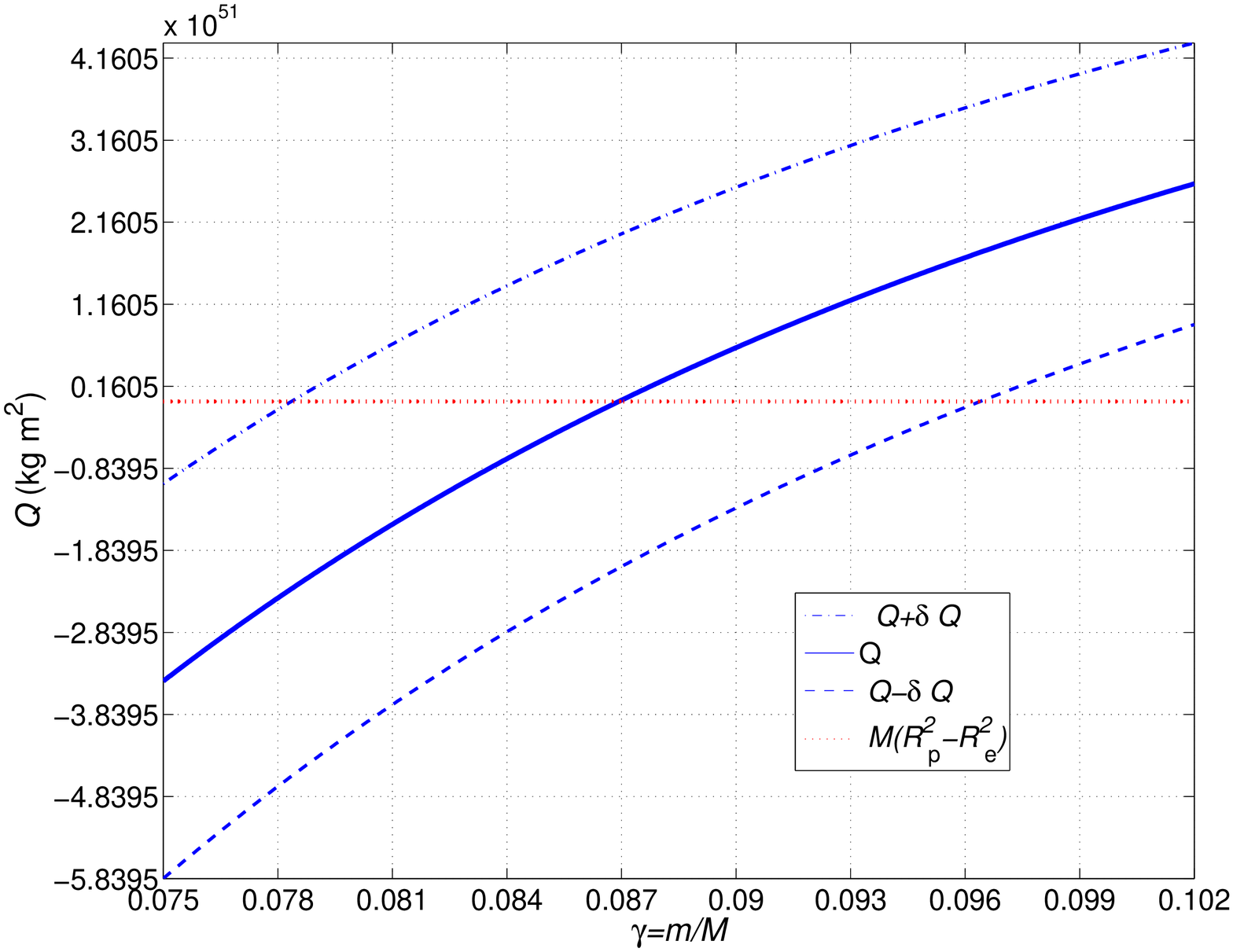}
   \caption{Allowed values for the quadrupole mass moment $Q$ as a function of $\gamma=m/M$ for $M=3.4$M$_{\odot}$ and $i=90$ deg.  For $\gamma\gtrsim 0.096$ $Q$ becomes positive, i.e. the star would be prolate, contrary to the observational evidence. For $\gamma\lesssim 0.078$ $Q$ is negative, but one-two orders of magnitude larger than $M(R_{\rm p}^2-R_{\rm e}^2)$.}
   \label{regulus_Q}
   \end{figure}
    %
    %
    %
 %
%
    %
    %
    %
Figure \ref{regulus_Q} clearly shows that the answer is negative: indeed, for  $\gamma \gtrsim 0.096$ the quadrupole mass moment $Q$ of the primary would become positive, i.e. the star would be prolate, which is contradicted by the observations. Thus, we can conclude that the faint companion  of Regulus cannot have a mass larger than about 0.30  M$_{\odot}$, contrary to what can be found in \citep{Gies08}.

An inspection of Figure  \ref{regulus_Q} tell us that for $\gamma \lesssim 0.078$, i.e. $m\lesssim 0.26$M$_{\odot}$, $Q$ would be a definite negative quantity, in according with the observations. However, it must be noted that, in this case, $|Q|$ would be one-two orders of magnitude larger than the estimate of \rfr{inerz}. It would be difficult to consider such a large discrepancy as acceptable. Thus, we provisionally consider $\gamma\lesssim 0.078$ not likely.

The simplest solution consists in assuming for $\gamma $ the range which still makes $\Delta P$ compatible with zero considering also that the value of  \rfr{inerz} for $Q$ would not be excluded  by such  values of $\gamma$, as shown by Figure \ref{regulus_Q}.

\section{Conclusions}
In this paper we investigated the impact of the huge quadrupole mass moment $Q$ of Regulus on the orbital period of its faint companion.

It turns out that, for a reasonable estimate of its value, $Q$ induces a correction to the Keplerian period which could, in principle, be measured because it is larger than the error in phenomenologically measuring the orbital period $P_{\rm b}$. However, according to the present-day level of knowledge of the system's parameters, the uncertainty in the Keplerian period is larger than the correction due to $Q$ by about two orders of magnitude; the most important sources of errors are the velocity semiamplitude $K$ and the mass $M$ of Regulus.

An analysis of deviations from the third Kepler law as due to $Q$ for the edge-on orbital configuration and $M=3.4(2)$M$_{\odot}$ has shown that values of the ratio $\gamma$ of the secondary's mass to the primary's one larger than about 0.096 are to be ruled out because they would yield a positive quadrupole mass moment for Regulus, i.e. it would be prolate, contrary to the observations. On the other hand, for $\gamma\lesssim 0.078$ $Q$ would be negative, but with an unlikely large value.

Further observational campaigns with different techniques would be of great significance in the attempt of reducing the overall uncertainty in the system's parameters down to the level required for detecting the effect of the shape of Regulus.


\end{document}